\documentclass{article}

\usepackage{PRIMEarxiv}

\usepackage[utf8]{inputenc} % allow utf-8 input
\usepackage[T1]{fontenc}    % use 8-bit T1 fonts
\usepackage{hyperref}       % hyperlinks
\usepackage{url}            % simple URL typesetting
\usepackage{booktabs}       % professional-quality tables
\usepackage{amsmath,amssymb}
\usepackage{amsfonts}       % blackboard math symbols
\usepackage{nicefrac}       % compact symbols for 1/2, etc.
\usepackage{microtype}      % microtypography
\usepackage{lipsum}
\usepackage{fancyhdr}       % header
\usepackage{graphicx}       % graphics
\graphicspath{{media/}}     % organize your images and other figures under media/ folder

\title{A continuous beam monochromator for matter waves
}

\author{
  Johannes Fiedler, Bodil Holst \\
  Department of Physics and Technology\\ University of Bergen\\All\'egaten 55, 5007 Bergen, Norway\\
  \texttt{johannes.fiedler@uib.no} \\
}

\begin{document}
\maketitle

\begin{abstract}
Atom and, of late, molecule interferometers find application in both the crucible of fundamental research and industrial pursuits. A prevalent methodology in the construction of atom interferometers involves the utilisation of gratings fashioned from laser beams. While this approach imparts commendable precision, it is hampered by its incapacity to attain exceedingly short wavelengths and its dependence on intricate laser systems for operational efficacy. All applications require the control of matter waves, particularly the particle's velocity. In this manuscript, we propose a continuous beam monochromator scheme reaching enormously high velocity purification with speed ratios in the order of $10^3$ based on atom-surface diffraction. Beyond these high purifications, the proposed scheme simplifies the application by reducing the degree of freedom to a single angle, selecting the wanted particle's velocity.
\end{abstract}

% keywords can be removed

\section{Introduction}
\label{intro}
Atom interferometry is one of the advanced investigation techniques in modern physics\cite{AtomInterferometry,D2CP03349F} covering a wide range from fundamental research, such as the transition between the classical and the quantum world due to high mass~\cite{Fein2019} or slower particles,~\cite{Taillandier-Loize_2016,PhysRevLett.127.170402} via as well as magnetic and gravity sensing~\cite{stray2022quantum,hardman2016simultaneous}, quantum metrology~\cite{riedel2010atom-chip-based}, atomic clocks~\cite{ludlow2015optical}, dark matter and gravitational wave detectors~\cite{canuel2020elgar} also in space~\cite{el-neaj2020aedge,Tino2013} to matter-wave lithography.~\cite{PhysRevApplied.11.024009,Fiedler_2022} Recently, portable atom gravimeters for geophysical investigations, such as prospecting and oil survey, have become commercially available.~\cite{desruelle2018} 

Atom interferometers are also proposed as accelerometers for sub-sea navigation in submarines and underwater drones~\cite{app10041256,hardman2016simultaneous}. However, reaching the envisaged accuracy requires either a velocity-sensitive measurement or an accurate velocity preselection.  Velocity-sensi\-tive measurements are challenging but realisable.~\cite{PhysRevLett.127.170402,Juffmann2012,Brand2015} 

A measure for the wave's monochromaticity is speed ratios (ratio between velocity $v$ and velocity spread $\Delta v$, $v/\Delta v$). 
To reach large speed ratios, one needs control of the particle's trajectories\cite{Juffmann2012,Arndt1999} and a low particle flux\cite{PhysRevLett.127.170402} to distinguish each particle. An alternative to a velocity-sensitive measurement is velocity preselection, enabling high-contrast interferences. Here, we differentiate between two principles: (a) changing the momentum of particles with the wrong velocity (momentum) or (b) removing the particles with a different momentum.  

Velocity-dependent accelerating or decelerating particles within a particle beam have been realised by the Rydberg--Stark decelerator, a chain of quadrupoles creating an inhomogeneous field that couples differently to the dipoles moving with different velocities.\cite{Hogan2016,VandeMeerakker2006,Tokunaga2009,VandeMeerakker2006} Thus, this technique is restricted to particles with a permanent dipole moment. Another possibility is the Zeeman slower, which works analogously to the Stark effect but uses magnetic fields coupled to the spin-polarised magnetic moment.

The more straightforward solution is removing the particles from the beam with a velocity different from the target velocity, which can be realised using two choppers, which only transmit particles with a velocity matching the time window for the chopper openings. This technique has been realised in various configurations and setups, e.g., a cascade of choppers~\cite{10.1063/1.5044203} or a helical gearwheel\cite{Juffmann2009,10.1063/1.3499254}. However, it has the apparent disadvantage that a continuous beam will be pulsed. A further possibility, removing particles with an unwanted velocity from a beam, uses atomic mirrors, which has been demonstrated experimentally\cite{Holst1997} as well as the slowing of atomic beams, the so-called atomic paddle~\cite{PhysRevLett.98.103201}, Stark effect decelerator~\cite{PhysRevA.82.013410} and Zeeman slower~\cite{PhysRevLett.48.596}.

Here, we present a novel approach to velocity selection which enables a continuous beam with speed ratios up to several hundred by exploiting the recently proposed reflective atom interferometer~\cite{Fiedler23}. The example presented here is for a helium beam scattering off hydrogen-passivated Si(111).\cite{doi:10.1063/1.480723} However, the proposed device can be adapted to other materials and atomic beams.  The reflection direction is velocity-sensitive depending on the surface structure. Thus, by sending the reflected (diffracted) beam through a pinhole, the particles with a velocity outside a specific range will be blocked, and the beam's speed ratio will be enhanced. The velocity-dependent beam spread is increased using three reflections instead of one (simple reflection scheme). Experimentally, this is made possible by the monolithic nature of the atom interferometer, which ensures that the reflective surfaces do not move relative to each other. 

\section{The monochromator}

\begin{figure}[htb]
    \centering
    \includegraphics[width=0.4\columnwidth]{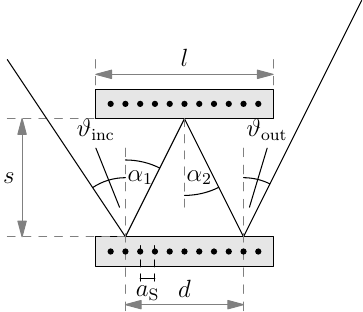}
    \caption{Sketch of the setup: two parallel nano-structured slabs of length $l$ and period $a_{\rm S}$ separated by the distance $s$. An incoming beam with incidence angle $\vartheta_{\rm inc}$ is reflected three times: first into the diffraction order $n_1$ with diffraction angle $\alpha_1$, followed by the diffraction order and angle $n_2$ and $\alpha_2$, respectively, and leaving the apparatus after a third reflection of order $n_3$ and the diffraction angle $\vartheta_{\rm out}$. The distance between the first and third reflection points is marked by $d$.}
    \label{fig:setup}
\end{figure}

The proposed monochromator for matter waves is based on the reflective atom interferometer introduced in Ref.~\cite{Fiedler23}, which consists of two parallel structured plates cut into a single crystal. It requires two parallel nano-structured planar surfaces, which can be achieved by cutting a monolithic crystal, such as silicon, and chemically dipping the Si(111) crystal in an HF solution.\cite{MACLAREN2001285} The incoming beam is diffracted three times within the device before leaving. For each diffraction, the incidence $\vartheta_{\rm inc}$ and reflection angles $\vartheta_{\rm out}$ are related by Bragg's law
\begin{equation}
    \vartheta_{\rm out} = \arcsin\left[ \sin\vartheta_{\rm inc} +n\frac{\lambda_{\rm dB}}{a_{\rm S}}\right]\,,\label{eq:bragg}
\end{equation}
with the diffraction order $n$, the de-Broglie wavelength $\lambda_{\rm dB} = 2\pi\hbar/p=2\pi\hbar/(mv)$, the reduced Planck constant $\hbar$, the particle's mass $m$ and velocity $v$, and the lattice constant of the structured surface $a_{\rm S}$. Due to the structure of this equation, the three internal diffractions $n_i$ lead to a total diffraction order for the entire device $N=n_1+n_2+n_3$. Figure~\ref{fig:setup} illustrates the situation.
In previous work~\cite{Fiedler23}, we have seen a strong dependency of the diffraction orders' positions on the incoming beam's wavelength, which motivated the further consideration of the device to act as a monochromator. In the following, we will first consider the relation between the incidence and reflection angle with respect to velocity deviations. One of these angles should be fixed for practical applications, where we used the outgoing angle $\vartheta_{\rm out}$. Thus, we first analyse the optimal incidence angle concerning a strong velocity dependence. Afterwards, we derive conditions for the device's dimensions, expressed in the length-to-separation ratio $d/s$, allowing for an almost arbitrary scaling of the device. All given examples consider a helium beam in a monolithic Si(111)-H(1$\times$1) device according to Ref.~\cite{Fiedler23}.

\subsection{Optimal incidence angle}
The diffraction angles $\varphi$ in the monolithic monochromator are determined by\cite{Fiedler23} 
\begin{equation}
    \vartheta_{\rm out} = \arcsin\left[\sin\vartheta_{\rm inc} + N\frac{2\pi\hbar}{mva_{\rm S}}\right]\,.
\end{equation}
%with the incidence angle $\vartheta_{\rm inc}$, the total diffraction order $N$, the reduced Planck constant $\hbar$, the particle's mass $m$ and velocity $v$ and the surface structure's period $a_{\rm S}$. 
To diffract most of the unwanted velocities of the incoming beam, a large variance of the diffraction angle with respect to velocity changes is required
\begin{align}
    \frac{\mathrm d \vartheta_{\rm out}}{\mathrm d v} =-\frac{N 2\pi\hbar}{ m v^2 a_{\rm S}\sqrt{1-\left(\sin(\vartheta_{\rm inc}) + N\frac{2\pi\hbar}{m v a_{\rm S}}\right)^2 }} \mapsto \rm{max.} \label{eq:dphi_dv}
\end{align}
Thus, the diffraction angle must be large to achieve a wide spread of the velocity distribution. The optimum would be mathematically at $\vartheta_{\rm inc}=\pi/2$. However, this solution means that the outcoming beam is parallel to the slab, leaving the application scope of Eq.~\ref{eq:bragg}. We introduce a small angle $\varepsilon$ to compensate for this effect, ensuring non-parallel beams, and the outcoming diffraction angle reads $\varphi = \frac{\pi}{2}-\varepsilon$ leading to an incidence angle
\begin{equation}
    \vartheta_{\rm inc} =\arcsin\left(\cos\varepsilon -N\frac{2\pi\hbar}{mva_{\rm S}}\right)\,.\label{eq:incindence}
\end{equation}
Furthermore, a small lattice constant is required $a_{\rm S} \mapsto 0$.

\begin{figure}[t]
    \centering
    \includegraphics[width=0.6\columnwidth]{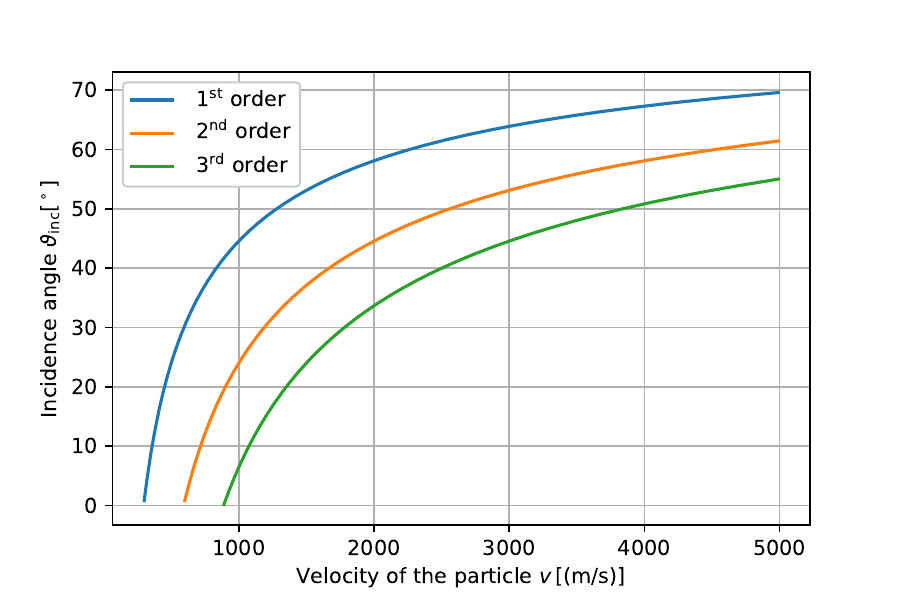}
    \caption{Incidence angles for a helium beam in a monolithic Si(111)-H(1$\times$1) monochromator for different velocities leading to a large diffraction angle ($\vartheta_{\rm out}=1.48 \,{\rm{rad}} \,(=85^\circ)$) for the first (blue line), second (orange line) and third (green line) total reflection order. It can be seen that the lowest diffraction order covers the largest velocity range (down $300\,\rm{m/s}$). The second order diffraction is bounded by $600\,\rm{m/s}$ and the third one by $890\,\rm{m/s}$.}
    \label{fig:incidence}
\end{figure}

\begin{figure}[t]
    \centering
    \includegraphics[width=0.6\columnwidth]{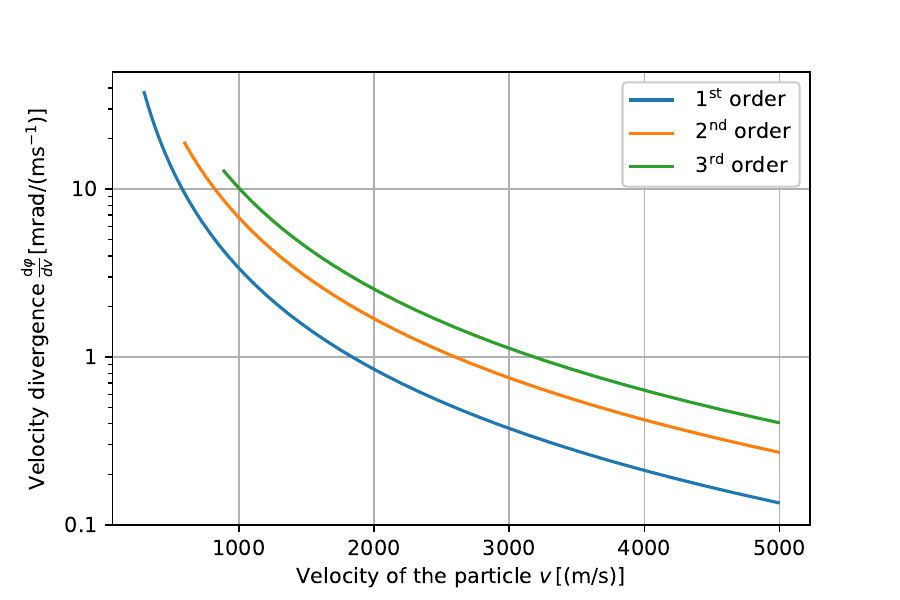}
    \caption{Velocity divergence for the diffraction of a helium beam for the incidence angles in Fig.~\ref{fig:incidence}.}
    \label{fig:dphi_dv}
\end{figure}

For applicational purposes, we set the output beam to a specific value, $\vartheta_{\rm out}=1.48 \,{\rm{rad}} \,(=85^\circ)$. This means the device always works with a fixed output angle, and the velocity selection occurs by changing the input angle. Figure~\ref{fig:incidence} illustrates the dependence of the incidence angles for different beam velocities of a helium beam diffracted in a monolithic hydrogen-passivised silicon monochromator with a lattice constant $a_{\rm S}=3.383$~\AA.~\cite{BARREDO200724} It can be seen that the lower diffraction orders cover a larger velocity range.
Figure~\ref{fig:dphi_dv} illustrates the corresponding velocity divergence~(\ref{eq:dphi_dv}). It can be seen that the higher diffraction orders are more sensitive to velocity changes than the lower ones. 

\subsection{Optimal dimensions of the monochromator}
\begin{figure}[t]
    \centering
    \includegraphics[width=0.6\columnwidth]{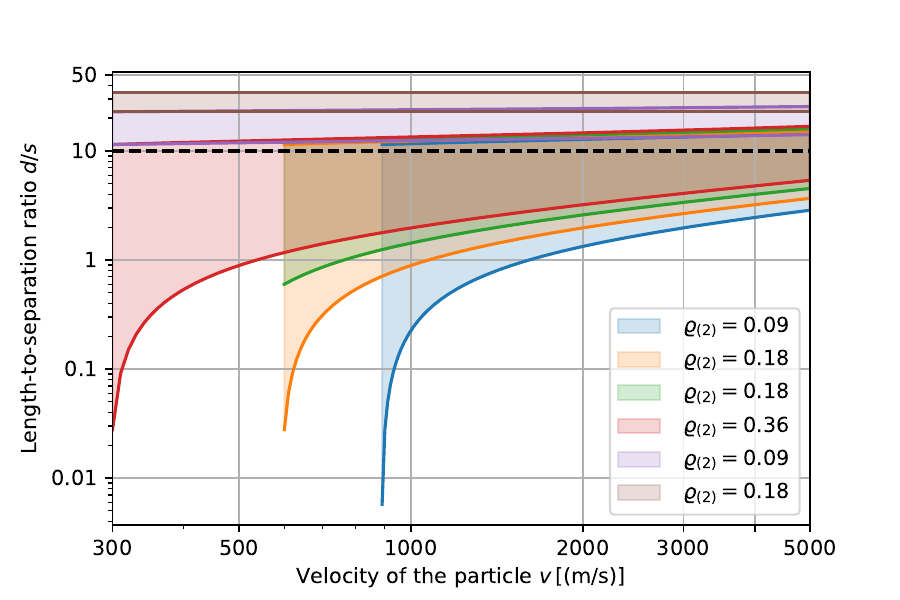}
    \caption{Dimension ranges for the monochromator operating on the first total diffraction order ($N=-1$) for a helium beam inside a monolithic Si(111)-H(1$\times$1) device for different velocities. The incidence angle is chosen according to Eq.~(\ref{eq:incindence}), $\vartheta_{\rm inc} = 85^\circ$, to achieve a large beam separation with respect to velocity deviations. Due to considering five diffraction orders for each reflection, 25 combinations of the internal diffraction orders $n_1$ and $n_2$ are possible, leading to sixteen paths grouped into twelve different paths. Six are depicted and labelled by the corresponding transmission rates $\varrho_{(2)} (\%)=a_1a_2$. The transparent area marks the range of the length-to-separation ratio $d/s$ bounded in the lower limit by the condition for three reflections and in the upper limit by leaving the device. The horizontal dashed black line marks the ratio $d/s=10$ used in further considerations.}
    \label{fig:lentosep}
\end{figure}

The monochromator is based on three diffractive reflections between two parallel plates separated by a distance $s$. The first reflection is described by the diffraction angle
\begin{equation}
    \alpha_1 = \arcsin\left[\sin\vartheta_{\rm inc} + n_1\frac{2\pi\hbar}{mva_{\rm S}}\right] \,,
\end{equation}
the second one by
\begin{equation}
    \alpha_2 = \arcsin\left[\sin\alpha_1 + n_2\frac{2\pi\hbar}{mva_{\rm S}}\right] \,,
\end{equation}
and finally, the exiting beam
\begin{equation}
    \vartheta_{\rm out} = \arcsin\left[\sin\alpha_2 + n_3\frac{2
    \pi\hbar}{mva_{\rm S}}\right]\,.
\end{equation}
By inserting these equations into each other, one finds the condition for the last diffraction order
\begin{equation}
    n_3 = N-n_1-n_2\,. 
\end{equation}
To determine the extension of the device, the diffractions need to occur within a certain length $d$ (spatial distance between reflection 1 and 3) smaller than the length of the device $l$, whose ratio to separation is given by the diffraction angles
\begin{equation}
    \frac{d}{s} = \tan\alpha_1 +\tan\alpha_2\,,
\end{equation}
which is bounded by
\begin{align}
   &d <l<d+ s\tan\varphi \\
   \Leftrightarrow & \tan\alpha_1 +\tan\alpha_2 <\frac{l}{s} < \tan\alpha_1 +\tan\alpha_2 +\tan\varphi\,. \label{eq:range}
\end{align}
The lower bound ensures three reflections and the upper bound for the beam leaving the device. Figure~\ref{fig:lentosep} illustrates this condition for a monochromator made off monolithic Si(111)-H(1$\times$1) for a helium beam operating on the first total diffraction order $N=-1$. We consider the internal diffraction orders to be from -2 to 2. Thus, 25 paths are possible, from which sixteen occur due to the criterion~\eqref{eq:range}, and eight are pairwise equal, leading to twelve different paths. In Figure~\ref{fig:lentosep}, we depicted six paths, which are labelled by the total transmission rate $\varrho$ as the product of the diffraction populations at each diffraction point, $\varrho_{(3)} = a_1 a_2 a_3$ (index $(3)$ labels the number of considered reflections), where we used the reflection probability for the zeroth, first and second order diffraction $a_i\in \left\lbrace 0.06, 0.03, 0.015\right\rbrace$ according to Ref.~\cite{Fiedler23,BARREDO200724}. It can be observed that due to the large diffraction angle for the outgoing beam ($\tan\varphi$), a wide range of length-to-separation ratios can be covered. Furthermore, not all combinations of internal diffraction orders fit all particle velocities. A ratio of 10 will cover the entire considered velocity range. 

\subsection{Continous velocity selection - a practical example}
\begin{figure}[t]
    \centering
    \includegraphics[width=0.6\columnwidth]{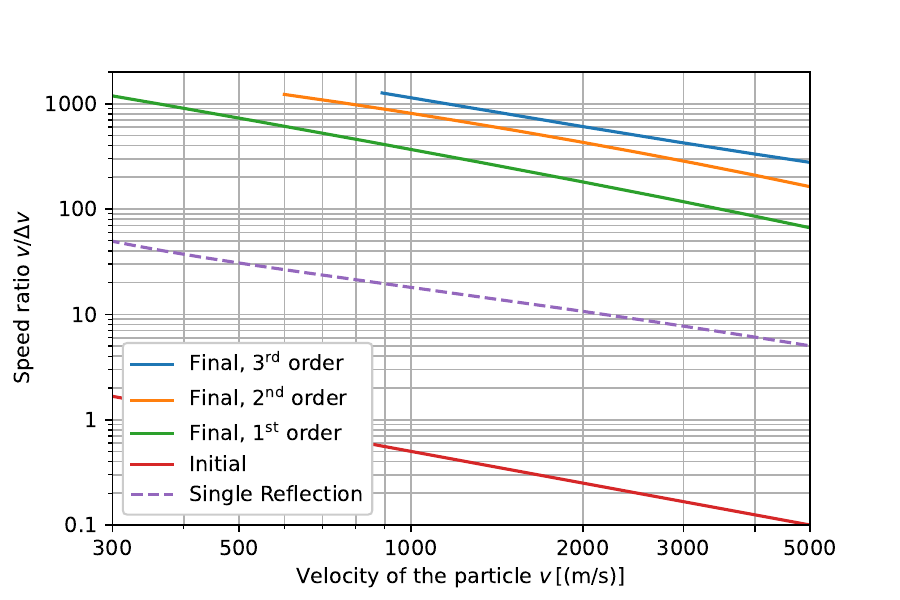}
    \caption{Comparison of the initial and final speed ratio for a helium beam passing the monolithic Si(111)-H(1$\times$1) monochromator.}
    \label{fig:speed}
\end{figure}

Let us consider a monolithic Si(111)-H(1$\times$1) monochromator with lattice constant $a_{\rm S}=3.383$~\AA,\cite{BARREDO200724} a length $d=5\,\rm{cm}$ and a slap separation $s=5\,\rm{mm}$. These parameters lead to a length-to-separation ratio of 10, covering the entire velocity range from 300 to 5000 m/s for the helium beam, see Fig.~\ref{fig:lentosep}. The lower bound is determined by the period of the slab's structure $a_{\rm S}$. Another material has to be chosen to reach lower velocities. We consider a plane wave entering the device through a pinhole with a 1 mm diameter 10 cm away from the device under an optimal incidence angle $\vartheta_{\rm inc}(v)$ according to Eq.~(\ref{eq:incindence}) for a total reflection angle of $85^\circ$. We have chosen the first total diffraction order as the working point, blue lines in Figs~\ref{fig:incidence} and \ref{fig:dphi_dv}. In addition, we considered two further pinholes 50 cm and 1 m after the device with a diameter of 1 cm each to block the background signal, such as contributions from other diffractions. For the simulation, we considered a rectangular velocity distribution with a symmetric width of 500 m/s around the central velocity, $v=\overline{v}\pm 250\,\rm{m/s}$. We calculated the propagation of this wave package using ray optics and determined the speed ratio $v/\Delta v$ after passing the final pinhole. The results are depicted in Fig.~\ref{fig:speed}. It can be seen that the monochromator yields excellent speed ratios for small particle velocities (below $1000 \,\rm{ms}^{-1}$) where it reaches values around 1000. In the medium range, around 2000 m/s, it is comparable to and slightly better than current techniques. The speed ratio decreases for high particle velocities due to the worse velocity divergence in this range; see Fig.~\ref{fig:dphi_dv}. The speed ratio for a single reflection with an incidence angle of $50^\circ$, achieving a transmission over a wide velocity range into the first order, is compared. It can be seen that the triple reflection monochromator yields an enhancement by one order of magnitude.

\subsection{Discussion}
We introduce a novel monochromator scheme for matter waves and demonstrate the resulting large achievable speed ratios (up to 600 to 1000) over a wide velocity range (up to $2000\,\rm{ms}^{-1}$). For a numerical example, we restricted ourselves to the consideration of helium beams. However, the device can be adapted to several other species. An increase in the particle's mass would yield a left shift of all curves (Figs.~\ref{fig:incidence}--\ref{fig:speed}) towards lower velocities. To adapt the monochromator to even smaller velocities, the lattice constant needs to increase to diffract the particles with lower velocities.

According to the incoming matter wave, its velocity spread must be restricted. It can be observed in Fig.~\ref{fig:incidence} that an incoming beam with a certain incidence angle will be diffracted to the same spot for three different velocities. The transmitted velocities have an integer ratio, $v_N = |N| v_1$, where $N$ denotes the total diffraction order, meaning that $v_2 =2v_1$ is transmitted via the second diffraction order and $v_3=3v_1$ via the third one, and so on. Thus, if the velocity spread of the incoming wave ($v/\Delta v < 1$) is wide enough to cover two or three total diffraction orders (Fig.~\ref{fig:incidence}), the outgoing beam will have a narrow velocity distribution around each of the two or three selected velocities.

The proposed monochromator uses three reflections to separate the velocity contributions of the beam. However, a single reflection would also reduce the velocity spread and provide a large transmission.~\cite{Holst1997} By comparing these different setups, there are some advantages for the three-reflection monochromator: (i) Due to the free-propagation between the reflection points, the velocity-dependent beam spread is wider, purifying the beam more than with a single reflection. (ii) The arrangement is easy to handle when the selected velocity is changed as it keeps one angle fixed.

\section{Conclusions}
We present a novel continuous beam monochromator scheme based on reflection atom interferometry, providing atom beams with high speed ratios and working across a broad velocity range. Due to the monolithic configuration, the proposed device will be easy to handle because only the incidence angle has to be tuned according to the wanted mean velocity. The theoretical calculations result in low transmission rates, which can be improved, for instance, by using quantum reflection~\cite{D2CP02641D} or reflection at evanescent potentials~\cite{PhysRevX.4.011029}.

\section*{Acknowledgments}
J.F. gratefully acknowledges support from the European Union (H2020-MSCA-IF-2020, grant number: 101031712).

%Bibliography
\bibliographystyle{unsrt}  
\bibliography{references}

\end{document}